\documentclass[allclo]{FBSart}
\usepackage{amsfonts}
\usepackage{amssymb}
\usepackage{epsfig}

\title{Point-Form Quantum Field Theory and Meson Form Factors}
\author{E.~P. Biernat\instnr{1}, K. Fuchsberger\instnr{1},
W.~H. Klink\instnr{2}, W.
Schweiger\instnr{1}\thanks{\textit{E-mail address:}
wolfgang.schweiger@uni-graz.at}}
\instlist{Institut f\"ur Physik, Universit\"at Graz, A-8010 Graz,
Austria \and Department of Physics and Astronomy, University of
Iowa, Iowa City, Iowa 52242, USA}

\runningauthor{W. Schweiger} \runningtitle{Point-Form Quantum
Field Theory} 

\begin{document}

\maketitle


\noindent Recently we have reconsidered the quantization of
relativistic field theories on a Lorentz-invariant surface of the
form $x_\mu x^\mu =\tau^2$~\cite{BKSZ07}. With this choice of the
quantization surface all components of the 4-momentum operator
become interaction dependent, whereas the generators of Lorentz
transformations stay free of interactions -- a feature
characteristic for Dirac's \lq\lq point form\rq\rq\ of
relativistic dynamics. Thus we speak of \lq\lq point-form quantum
field theory\rq\rq\ (PFQFT). Old papers on PFQFT (see,
e.g.,~\cite{S74,GRS74}) dealt mainly with the evolution of quantum
fields in the parameter $\tau$ and made use of a Fock-space basis
which is related to the generators of the Lorentz group. Such a
choice for the basis and the \lq\lq time parameter\rq\rq, however,
gave rise to conceptual difficulties. To avoid these problems we
have kept the usual momentum basis and considered evolution of the
system as generated by the 4-momentum operator~\cite{BKSZ07}. In
this way we were able to show for free fields that quantization on
the space-time hyperboloid $x_\mu x^\mu =\tau^2$ leads to the same
Fock-space representation of the Poincar\'e generators as
equal-time quantization. Moreover, we have suggested a generalized
interaction picture which leads to a manifestly Lorentz covariant
expression for the scattering operator as path-ordered exponential
of the interaction part of the 4-momentum operator (along
arbitrary timelike paths). We furthermore showed that the
perturbative expansion of the scattering operator, defined in such
a way, is (order by order) equivalent to usual time-ordered
perturbation theory.


The nice feature that the operator formalism becomes manifestly
Lorentz covariant if fields are quantized on the space-time
hyperboloid $x_\mu x^\mu =\tau^2$ was not our only motivation to
study PFQFT. PFQFT serves also as a natural starting point for the
construction of effective interactions, currents, etc., which can
be applied to point-form quantum mechanics. The main difficulty of
finding a quantum mechanical realization of the Poincar\'e
algebra, which describes a finite number of interacting particles,
is caused by the fact that interaction terms in the Poincar\'e
generators have to satisfy non-linear constraints, in
general.\footnote{These constraints are automatically satisfied if
a local interacting field theory is quantized.} A procedure that
resolves this problem has been proposed by Bakamjian and
Thomas~\cite{BT53}. Its point-form version amounts to the
assumption that the free 4-velocity operator
$\hat{V}^\mu_\mathrm{free}$ can be factored out of the
(interacting) 4-momentum operator
\begin{equation}
\hat{P}^\mu = \hat{P}^\mu_\mathrm{free} +\hat{P}^\mu_\mathrm{int}
=
(\hat{M}_\mathrm{free}+\hat{M}_\mathrm{int})\hat{V}^\mu_\mathrm{free}\,
,
\end{equation}
so that one is left with an interacting mass operator
$\hat{M}=\hat{M}_\mathrm{free}+\hat{M}_\mathrm{int}$. Since the
mass operator is a Casimir operator of the Poincar\'e group the
interaction term $\hat{M}_\mathrm{int}$ is then only restricted by
linear constraints, which are easy to satisfy.

As an example of how field theoretical concepts may enter the
framework of relativistic quantum mechanics let us consider the
electromagnetic form factor of a confined quark-antiquark pair
(e.g. the pion). The idea is to  work within the Bakamjian-Thomas
framework and treat the electromagnetic scattering of an electron
by a meson with internal structure as a quantum mechanical
2-channel problem for the mass operator, in which the dynamics of
the exchanged photon is explicitly taken into account. The
structure of the meson is encoded in a phenomenological vertex
form factor which is not known a priori. Similarly we can consider
the electromagnetic scattering of an electron by a quark-antiquark
pair, which interacts via an instantaneous confining potential, as
a 2-channel problem. If we reduce both 2-channel problems to
1-channel problems  for the $\mathrm{e}\mathrm{M}$ and
$\mathrm{e}\mathrm{q}\bar{\mathrm{q}}$ channels, respectively, we
end up with 1-photon exchange optical potentials for the two
systems (cf. Fig.~1). By comparing appropriate matrix elements of
the $\mathrm{e}\mathrm{q}\bar{\mathrm{q}}$ optical potential (i.e.
between states in which $\mathrm{q}$ and $\bar{\mathrm{q}}$ are
bound with the quantum numbers of the meson) with those of the
$\mathrm{e}\mathrm{M}$ optical potential the electromagnetic form
factor can be identified. The only problem with this kind of
procedure is that a simple factorization as in Eq.~(1) does not
hold for the full electromagnetic vertex. We therefore have to
resort to the approximation that the total 4-velocity of the
system is conserved at electromagnetic vertices. In
Ref.~\cite{K03} it has been demonstrated in some detail that this
is a way to implement field theoretical vertex interactions into a
Bakamjian-Thomas type framework.

\begin{figure}[h!]
\begin{center}
\epsfig{file=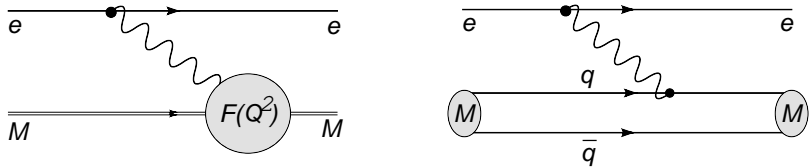,height=2.4cm,angle=0,clip=0}\hspace{1cm}
\vspace{-0.5cm} \caption{Contributions to the one-photon-exchange
optical potential for electron-meson scattering (left) and
electron scattering off a $\mathrm{q}\bar{\mathrm{q}}$-cluster
(right).}
\end{center}
\end{figure}

A calculation of the electromagnetic pion form
factor along the lines just sketched has been carried out in
Ref.~\cite{F07}. For simplicity quarks have been treated as
spinless. What one observes is that the extracted form factor
depends not only on the virtuality $-Q^2$ of the photon, but also
on the momentum of the meson $|\vec{k}_\mathrm{M}|$ in the
center-of-mass of the electron-meson system (Fig.~2 left). This is
equivalent to a dependence on Mandelstam $s$ and does not spoil
Poincar\'e invariance. It is merely a consequence of the
assumption that the 4-velocity of the system is conserved at
electromagnetic vertices. The $|\vec{k}_\mathrm{M}|$-dependence
vanishes rather quickly and if one takes the limit
$|\vec{k}_\mathrm{M}|\rightarrow \infty$ the optical potential
assumes its expected structure $V_{\mathrm{opt}}\propto
j_{\mathrm{e} \mu} j^\mu_{\mathrm{M}} / Q^2$ with
$j^\mu_{\mathrm{M}} = (k_{\mathrm{M}}+k_{\mathrm{M}}^\prime)^\mu
F(Q^2)$. In this limit the expression for the form factor becomes
\begin{equation}
F(Q^2=\vec{q}^2)=\int_{{\mathbb R}^3}
d^3\tilde{k}^\prime_\mathrm{q}
\sqrt{\frac{m_{\mathrm{q}\bar{\mathrm{q}}}}
{m_{\mathrm{q}\bar{\mathrm{q}}}^\prime}} \psi^\ast (
\vec{\tilde{k}}_\mathrm{q}^{\prime}) \psi (
\vec{\tilde{k}}_\mathrm{q})\,,
\end{equation}
where $\vec{k}_{\mathrm{M}}=\vec{k}_{\mathrm{M}}^\prime -
\vec{q}$, $\vec{k}_\mathrm{q} = \vec{k}_\mathrm{q}^\prime -
\vec{q}$ and
$m_{\mathrm{q}\bar{\mathrm{q}}}^2=(E_{\mathrm{q}}+E_{{\bar{\mathrm{q}}}})^2
- \vec{k}_\mathrm{M}^2$. Quantities without a tilde refer to the
electron-meson center-of-mass and quantities with a tilde to the
meson rest system. With a simple harmonic-oscillator wave function
$\psi (\, \vec{\tilde{k}}_\mathrm{q})$ and a reasonable choice of
the two free parameters (the constituent-quark mass $m_\mathrm{q}$
and the oscillator parameter $a$) a satisfactory fit of the pion
form-factor data up to momentum tranfers of a few GeV$^2$ can be
achieved (Fig.~2 right). There is, of course, room left for other
dynamical ingredients.

\begin{figure}[t!]
\epsfig{file=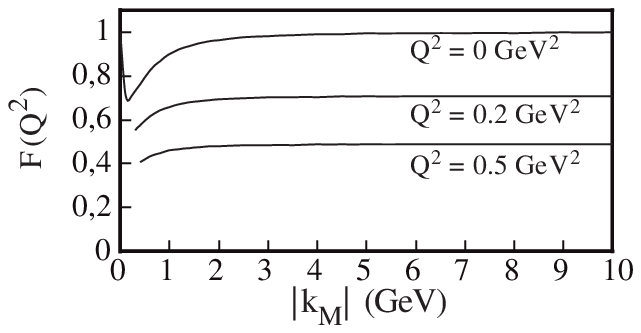,width=5.5cm,angle=0,clip=0}\hspace{1.5cm}
\epsfig{file=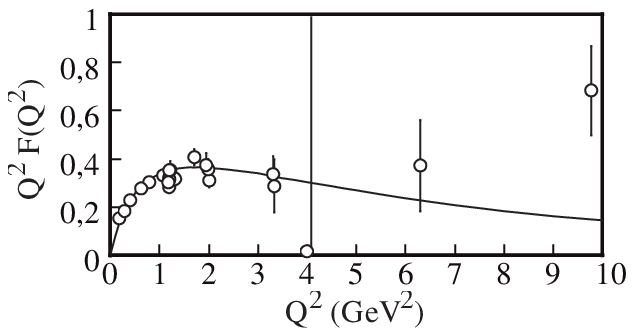,width=5.5cm,angle=0,clip=0}
\caption{Electromagnetic pion form factor calculated with a
harmonic-oscillator wave function $\propto
\exp(-\vec{\tilde{k}}_\mathrm{q}^2/2 a^2 )$($a=0.21$~GeV,
$m_\mathrm{q}=0.2$~GeV). The dependence on the meson CM momentum
$|\vec{k}_\mathrm{M}|$ (momentum transfer $Q^2$ fixed) is shown in
the left plot, whereas the $Q^2$ dependence for
$|\vec{k}_\mathrm{M}| \rightarrow \infty$ (cf. Eq.~(2)) is
depicted in the right plot. Data are taken from Ref.~\cite{B78}.}
\end{figure}

The big advantage of our approach is that the electromagnetic
current for a hadron with internal structure comes out with the
right properties since we work within a manifestly Lorentz
covariant framework. These first exploratory calculations for a
very simple bound system show the virtues of point-form dynamics.
The next step will be the inclusion of the quark spin.
Generalizations to vector mesons, (spin--1/2 and spin--3/2)
baryons and hadron transition form factors seem to be quite
obvious and will be the focus of our further investigations.


\end{document}